# Absolute measurement of thermal noise in a resonant short-range force experiment

H Yan[1], E A Housworth[2], H O Meyer[1], G Visser[3], E Weisman[1], and J C Long[1,4]

[1]Department of Physics, Indiana University, Bloomington, Indiana 47405, USA
[2]Department of Mathematics, Indiana University, Bloomington, Indiana 47405, USA
[3]Center for Exploration of Energy and Matter, Indiana University, Bloomington, Indiana 47408, USA

Planar, double-torsional oscillators are especially suitable for short-range macroscopic force search experiments, since they can be operated at the limit of instrumental thermal noise. As a study of this limit, we report a measurement of the noise kinetic energy of a polycrystalline tungsten oscillator in thermal equilibrium at room temperature. The fluctuations of the oscillator in a high-$Q$ torsional mode with a resonance frequency near 1 kHz are detected with capacitive transducers coupled to a sensitive differential amplifier. The electronic processing is calibrated by means of a known electrostatic force and input from a finite-element model. The measured average kinetic energy, $E_{exp} = (2.0 \pm 0.3) \cdot 10^{-21}$ J, is in agreement with the expected value of $\frac{1}{2} k_B T$.

PACS numbers: 46.80.+j, 05.40.-a, 04.80.Cc, 07.07.Df, 07.10.Fq

## 1 Introduction

The equipartition theorem predicts that any physical system in thermal equilibrium is associated with energy. In particular, this holds for mechanical systems used in precision force measurements, where the random thermal motion of the detector represents one of the dominant noise sources and may limit the sensitivity that can be achieved [1].

In the experiment reported here, we investigate thermal noise by carrying out a measurement of the random kinetic energy $E_{exp}$ of a torsional thin-plate detector. In contrast to a previous study of thermal noise in a single-crystal silicon oscillator of similar shape [2], we take a somewhat different approach in which a calibration procedure and a finite-element model are used to produce an *absolute* measurement of $E_{exp}$ which can be compared directly to the prediction of the equipartition theorem.

---

[4] Author to whom correspondence should be addressed: jcl@indiana.edu

Absolute measurement of thermal noise in a resonant short-range force experiment

The present measurement employs a planar, double-torsional tungsten oscillator as shown in Figure 1. The device is based on a proven design for single-crystal silicon oscillators, developed for cryogenic, condensed matter physics experiments [3,4]. Our oscillator is somewhat simpler, consisting of three thin rectangles, interconnected by narrow strips. It is mounted by clamping the B rectangle. Given its application in the search for weak, mass-coupled forces (section 2.1), the oscillator is fabricated from polycrystalline tungsten since, at room temperature, this material offers a good compromise between a low dissipation factor and high mass density.

The oscillator described here exhibits a number of resonances, or modes, with frequencies from ~100 Hz to many kHz. The mode of interest is the first 'anti-symmetric torsion' mode (AST) in which the F and M rectangles undergo torsional motion, opposite in phase, around the $y$-axis in figure 1. This mode is preferred because it has uniquely low intrinsic friction (high $Q$-value), as has been shown in a detailed study of the modes of a similar oscillator [5].

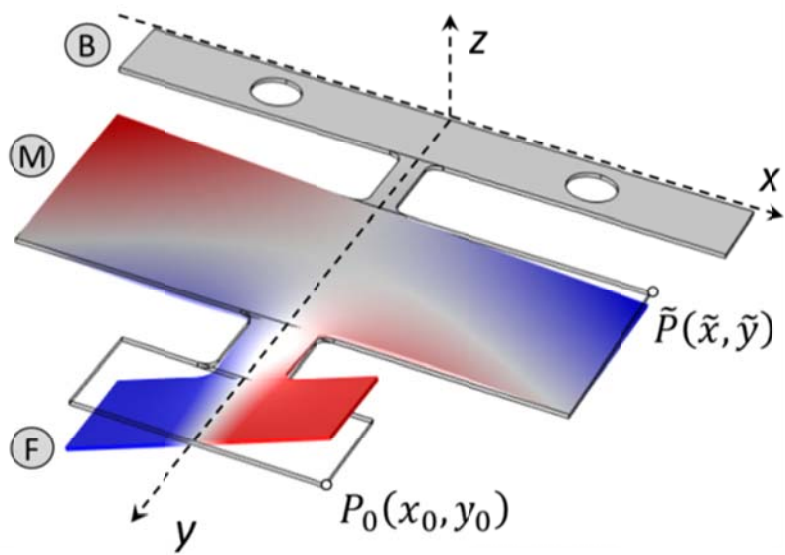

Figure 1. Detector motion corresponding to the AST mode. The two special points $\tilde{P}(\tilde{x},\tilde{y})$ and $P_0(x_0, y_0)$ are introduced in sections 3.1 and 4.3, respectively.

This paper is organized as follows. Section 2 contains a description of the apparatus. In section 3, the theoretical basis is outlined, and the quantities that are required for a determination of the thermal kinetic energy $E_{exp}$ are identified. The measurement or modeling of each of these quantities is presented in section 4. The final result and a discussion are presented in section 5.

## 2 Apparatus

### 2.1 *Experimental setup*

The oscillator described in the previous section is the 'detector' mass in an experiment to search for unobserved mass-coupled forces in the sub-millimeter range. Force search experiments (or, equivalently, tests of the gravitational inverse square law) in this range have received a great deal of attention in the past two decades. Present experimental limits allow for unobserved forces in nature several million times stronger than gravity acting over distances of a few microns, and predictions of unobserved forces in this range have arisen in several contexts, including attempts to describe gravity and the other fundamental interactions in the same theoretical framework. For comprehensive and recent reviews, see Refs. [6-9].

The experiment is designed for optimum sensitivity to mass-coupled forces in the range between 10 and 100 microns. The sub-millimeter range has been the subject of active theoretical investigation, notably on account of the prediction of "large" extra dimensions at this scale which could explain the hierarchy problem [10]. Many theories beyond the Standard Model possess extended symmetries that, when broken at high energy scales, lead to weakly-coupled, light bosons including moduli [11], dilatons [12], and axions [13], which can generate relatively long-range interactions. The fact that the dark energy density, of order $(1 \text{ meV})^4$, corresponds to a length scale of about 100 μm also encourages searches for unobserved phenomena at this scale [6].

Published limits on unobserved forces between this range and 1 cm are defined by classical gravity experiments. These include the short-range torsion pendulums at Irvine [14], the Huazhong University of Science and Technology (HUST) [15], and in the Eot-Wash group at the University of Washington [16]. Below 10 microns, the best limits derive from experiments using force microscopy and microelectromechanical techniques, including the Stanford microcantilever experiment [17], and from Casimir Force experiments [18, 19]. As explained in the conclusions, when used as a force search, the absolute sensitivity of the experiment described herein falls below the classical gravity experiments, however, greater sensitivity is possible at micron ranges.

The experiment is described in Ref. [20]; the essential details are reviewed here. The detector is designed to measure the force between it and a 'source' mass. The configuration of the two masses is shown in figure 2. The planar geometry is efficient in concentrating as much mass as possible at the range of interest, and in suppressing the $1/r^2$ term of the Newtonian gravitational force. The force-generating source mass is a flat beam fabricated from a 300 μm thick tungsten plate and mounted at one of the two nodes of its fundamental transverse mode. When the experiment is operated as a force search, the source mass is arranged so that it overlaps half of the detector F plate and is driven piezoelectrically at the

Absolute measurement of thermal noise in a resonant short-range force experiment frequency of the AST mode of the detector. A stiff conducting shield (not shown in figure 2) is placed between the source mass and the detector to eliminate electrostatic backgrounds caused by surface potentials and acoustic backgrounds conveyed by the residual gas. The original shield [20] consisted of a 60 micron thick sapphire plate with a 100 nm gold coating. It has been replaced in the present apparatus with a 10 micron thick stretched beryllium copper membrane of comparable compliance, which permits smaller test mass separation, and potentially much greater sensitivity to interactions in the micron range.

Detector oscillations are read out with a pair of capacitive transducers coupled to a custom, low-noise preamplifier (section 2.4.1). The probes are situated over the rear corners of the large detector rectangle, where they can be electrically shielded from the interaction region. The differential design, based on a pair of op-amps, replaces a single-ended design based on a discrete JFET [20] and provides increased protection against ground loops.

For the present measurement, which is the characterization of the limiting statistical background and thus of the absolute sensitivity of the current apparatus with the new readout, the source mass is at rest and the shield is omitted. Otherwise, the apparatus is the same as is used for the force search experiments.

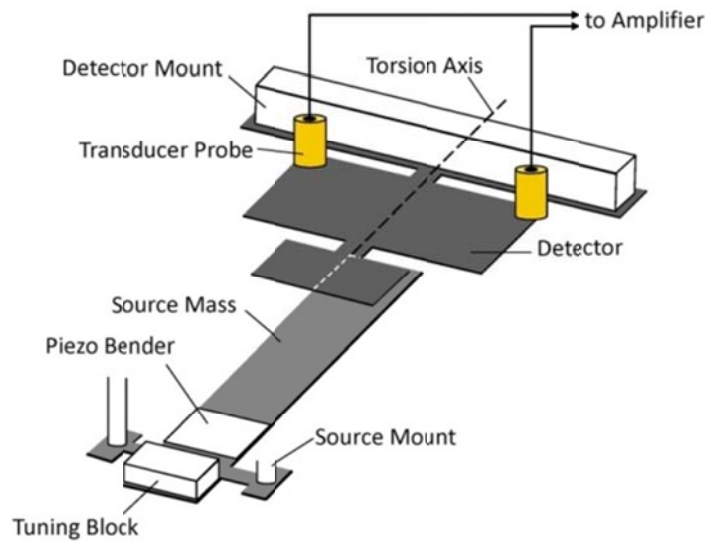

Figure 2. (adapted from [20]). Test mass configuration used for the thermal energy measurements, showing the location of the capacitive transducer probes. The source mass remains stationary for these measurements. When the experiment is operated as a short-range force search, the source is driven by the piezoelectric bender at the detector AST mode frequency, and a thin conducting shield is placed between the masses.

Absolute measurement of thermal noise in a resonant short-range force experiment

*2.2 Environment*

The resonant operation represents a challenge for the vibration isolation between the test masses. The detector oscillator has been designed such that its AST frequency is near 1 kHz, since in this frequency range it is possible to build a simple, passive vibration isolation system with a high degree of rigidity and dimensional stability, allowing the test masses to be maintained at very small separations for indefinite duration. This vibration isolation system consists of stacks of brass disks connected by very fine wires under tension, which form a series of soft springs relative to the 1 kHz operation frequency [21]. The detector and the source are suspended from independent stacks (not shown in figure 2), which are each in turn suspended from a three-axis micrometer stage allowing for rapid positioning of the test masses to within a few microns of precision. To further suppress acoustic backgrounds, the experiment is mounted in a 75-liter vacuum bell jar and the pressure is maintained in the $10^{-7}$ torr range with a turbomolecular pump.

The temperature of the setup is monitored with a silicon diode sensor attached to the detector mount. A heating resistor is used to keep the temperature at 302 K, a few degrees above ambient. The temperature is stable to well within 0.1 K.

The Newtonian, acoustic, and electromagnetic backgrounds are all associated with the resonant motion of the source. The vibration isolation system, shield, and vacuum were sufficient to eliminate any evidence of these resonant backgrounds in the experiment in [20], and given that the changes to the experiment are in the readout and a thinner shield with comparable compliance, we expect this to be the case for the current apparatus. When the resonant backgrounds are sufficiently controlled, the limiting background is expected to be the thermal noise of the detector oscillator. This was the case with the experiment in [20], but it could not be proven for the lack of an absolute calibration of the observed signal. This provided the main motivation for the research reported here.

*2.3 Preparation of the detector oscillator*

The oscillators are of the same design and composition as that used in [20], and prepared in a similar process. They are manufactured from high-purity tungsten plates of 200 μm thickness [22]. Prototype oscillators are cut into the form shown in figure 3 via electric-discharge machining [23].

Absolute measurement of thermal noise in a resonant short-range force experiment

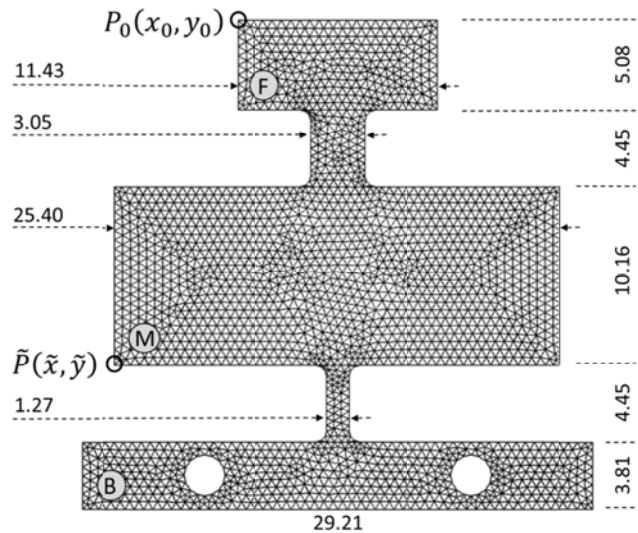

Figure 3. Detector geometry with front, middle and base plates (F, M, B). The dimensions are given in units of mm. The filet radius is 0.635 mm and the diameter of the mounting holes is 3.81 mm. Tolerances are better than 0.02 mm. The thickness of the plate is $d_w$ = 0.190 mm. The two special points $\tilde{P}(\tilde{x}, \tilde{y})$ and $P_0(x_0, y_0)$ are described in the text. Also shown is the meshing used in the finite-element simulation (section 4.3).

Ultrasonic cleaning [24] is used to eliminate most surface contamination. Burrs, other machining marks and remaining discoloration are subsequently removed with a 5 minute etch in modified Murakami reagent [25], followed by ultrasonic rinsing in distilled water. The etching reduces the average thickness by about 10 microns. High-temperature annealing to promote recrystallization has been shown to reduce the internal friction of metallic oscillators [26, 27]. The prototypes are annealed in vacuum ($P = 3 \cdot 10^{-5}$ torr) at 2573 K for two hours in an attempt to reach the secondary recrystallization temperature [28]. This is followed by slow cooling to room temperature. From a particular fabrication run, a prototype with high measured $Q$ and surface flatness is selected for noise measurements. The $Q$ of the particular prototype measured below, while among the highest available at the time, is slightly smaller than for the detector in Ref. [20].

## 2.4  Electronic processing

*2.4.1  Transducer and preamplifier.* The motion of the detector is observed by a capacitive pick-up, consisting of two circular electrodes of 4 mm diameter, parallel to the detector surface, located at the

Absolute measurement of thermal noise in a resonant short-range force experiment back-corners of the M rectangle near the reference point $\tilde{P}$ (see Figs. 1 - 3). The gap between the detector and the electrodes is about $z_p \sim 0.1$ mm, and the resulting capacitance is about 1 pF. A bias voltage $u_p$ is applied to the pick-up electrodes; the detector is at ground. The detector motion affects the capacitance and causes a current to flow, which is proportional to $u_p$ and the velocity of the detector surface.

For torsional motion, the currents from the two electrodes are opposite in phase. Taking advantage of this, a low-noise, differential amplifier has been designed, consisting of two symmetric, charge-sensitive, inverting amplifiers (figure 4). The circuit allows a probe bias of up to 500 V. The input current from the probes is capacitively coupled to the amplifier through an ultra-stable, ultra-low leakage current, monolithic glass capacitor. A gas discharge tube and 49.9 k$\Omega$ series resistor provide some protection for the op-amp in the event of breakdown in the sensor gap. The whole circuit, including mounts for the pick-up electrodes, is constructed on material of low dielectric constant and dielectric loss[5] to minimize its noise contribution. The operational amplifier[6] has been selected for low input bias current and low noise. The feedback capacitors are adjustable for common-mode rejection. The differential output voltage from the amplifiers is transformer-coupled to the output cable, to reject the common signal and to isolate the amplifier ground from the cable.

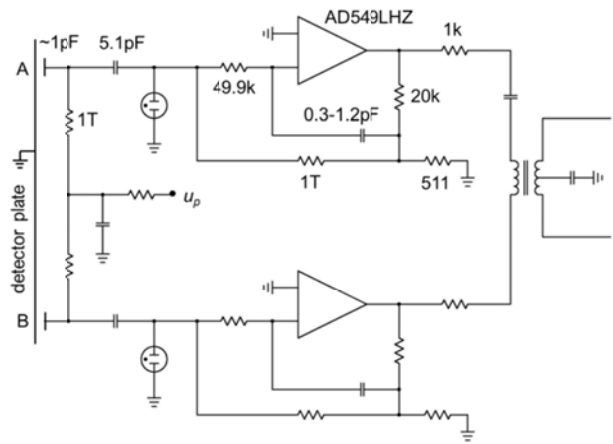

Figure 4. Pre-amplifier circuit. The pick-up electrodes A, B, facing the detector plate, are biased with voltage $u_p$. The circuit consists of two symmetric halves.

---

[5] RO-4003, Rogers Corp., $\varepsilon_R$ = 3.38, tan $\delta$ = 0.0021
[6] AD549LHZ, Analog Devices, www.analog.com

Absolute measurement of thermal noise in a resonant short-range force experiment

A measurement of the electronic noise, relative to the observed mechanical noise, is discussed in section 4.2.4.

*2.4.2 Remaining signal processing.* The preamplifier is followed by a second amplifier (Stanford Research SR560), and a two-phase lock-in amplifier (Stanford Research SR830, hereafter LIA). A crystal-controlled oscillator (Stanford research DS345) provides a reference frequency $f_{LIA}$ for the LIA. It is also used to drive the detector oscillator directly, in order to obtain a precise measurement of the mechanical quality factor ($Q$), and to apply an AC voltage to the source mass as part of the calibration procedure. Both measurements are necessary for absolute determination of the random noise. The second amplifier, operated in differential mode with low gain (typically 20) and rather strong high-pass filtering (typically 1 kHz cutoff and 12 dB/octave roll off), is used to avoid instabilities in the LIA differential input. Additional high-pass filter stages are added to the input of the second amplifier, since excitation of lower-order detector modes (presumably from room vibrations) can lead to strong transient signals that exceed the amplifier input rating.

The LIA provides an adjustable low-pass filter with an equivalent noise bandwidth $B_{LIA}$. This defines the measurement bandwidth $B = 2B_{LIA}$, centered at the reference frequency, $f_{LIA}$.

## 3   Definitions and theoretical outline

*3.1   Motion of the detector oscillator*

A normal mode of the detector oscillator is characterized by the sinusoidal, in-phase motion of all parts of the system, with a resonance frequency $\omega_0$ and with a specific mode shape. The AST mode of the detector oscillator is shown in figure 1. The displacement from equilibrium shown is exaggerated and, in reality, is small enough to assume that all motion is in the *z*-direction.

The displacement $z(x,y,t)$ has a sinusoidal dependence on time *t*. Its amplitude at a point $(x,y)$, arbitrarily normalized, is called the *mode shape function* $\xi(x,y)$. The mode shape function is obtained by modeling the oscillator within the framework of a finite-element calculation (section 4.3).

Once the mode shape is known, it is possible to express the displacement anywhere on the detector by the displacement $\tilde{z} \equiv z(\tilde{x}, \tilde{y})$ at an arbitrarily chosen 'reference' point $\tilde{P}$ by

$$z(x,y) = \tilde{z} \cdot \left\{ \xi(x,y) / \tilde{\xi} \right\} , \qquad (1)$$

Absolute measurement of thermal noise in a resonant short-range force experiment where $\tilde{\xi} \equiv \xi(\tilde{x}, \tilde{y})$. This makes it possible to describe the dynamics of a given, isolated mode of the oscillator by its displacement at a single point $\tilde{P}$. Thus, the system can be treated as a harmonic oscillator with a single degree of freedom, and with the equation of motion

$$\tilde{m}\ddot{\tilde{z}} + \gamma \dot{\tilde{z}} + k\tilde{z} = \tilde{F} \quad . \tag{2}$$

Here, $\tilde{F}$ is a periodic driving force of frequency $\omega = 2\pi f$, acting at point $\tilde{P}$ in the z-direction and $\tilde{z}$ is the system response at the same frequency. The term $\gamma\dot{\tilde{z}}$ represents a frictional force and $\tilde{m}$ is the so-called *modal mass*, which will be defined in section 4.3. The spring constant $k$ is related to the resonance frequency $\omega_0$ by $k = \tilde{m}\omega_0^2$.

The response function $\chi(\omega)$, defined by $\tilde{z} \equiv \chi(\omega) \cdot \tilde{F}$, is obtained using $\dot{\tilde{z}} = i\omega\tilde{z}$, $\ddot{\tilde{z}} = -\omega^2\tilde{z}$, and the approximation $\omega_0^2 - \omega^2 \approx 2\omega(\omega_0 - \omega)$, valid for a narrow resonance. Then,

$$\chi(\omega) = \frac{1/\omega}{2\tilde{m}(\omega_0 - \omega) + i\gamma} \quad . \tag{3}$$

Energy dissipation is often measured by the quality factor, or *Q*-value, which will be determined from the 1/e decay time $\tau$ of the amplitude of free oscillations in section 4.1. The *Q*-value is related to the friction parameter $\gamma$ and to the full width $\Delta\omega$ of $|\chi(\omega)|^2$ at half maximum by

$$Q = \tfrac{1}{2}\omega\,\tau = \frac{\tilde{m}\omega}{\gamma} = \frac{\omega}{\Delta\omega} \quad . \tag{4}$$

The displacement amplitude $\tilde{z}_D$ in response to a driving force with amplitude $\tilde{F}_D$ *at the resonance frequency* is now given by

$$\tilde{z} = |\chi(\omega)|\tilde{F} = \frac{Q\tilde{F}}{\tilde{m}\omega} \quad . \tag{5}$$

As explained in [1], the form of the equation of motion depends on the nature of the damping in the oscillator. Eq. (2) above assumes velocity damping from an external source such as viscous drag or a lossy mount. Internal damping is another possibility, which can be described in terms of an imaginary spring constant. Both internal and external sources appear to contribute to damping in the oscillator

Absolute measurement of thermal noise in a resonant short-range force experiment investigated below. The former is evidenced by the improvement in $Q$ after annealing, the latter by variability in $Q$ observed with different mounting schemes and clamping pressures. Since the noise measurements in this study are taken near resonance, however, the source of damping has no bearing on the analysis. Power spectra of an oscillator with the same relevant characteristics as the one measured here (from section 4: $\tilde{m} = 0.93$ g, $\omega_0/2\pi = 1223$ Hz, $Q = 20800$) can be computed for the case of external damping (Eq. (4) of [1]) and internal damping (Eq. (16) of [1]). The fractional difference between these spectra over the entire frequency range of the noise measurements ($\omega_0/2\pi \pm 2.5$ Hz) is never greater than $2\times10^{-3}$. For simplicity, an external damping term is assumed in Eq. (2).

*3.2 Thermal noise motion*

For any system in thermal equilibrium at temperature $T$, the principle of equipartition of energy states that the mean energy associated with each variable that contributes a quadratic term to the Hamiltonian, equals $\tfrac{1}{2}k_B T$. A harmonic oscillator has two such terms, one for potential and one for kinetic energy. Thus, we expect that the average kinetic energy due to random thermal fluctuations of the oscillator equals

$$\tfrac{1}{2}k_B T = 2.085 \cdot 10^{-21}\,\text{J} \quad (\text{at } T = 302\,\text{K}) \tag{6}$$

To maintain this motion in the presence of frictional losses, a driving force is needed. This force, supplied by the environment, is random in time and independent of frequency [18] with a mean-square (unilateral) spectral density $\langle F_{th}^2 \rangle'$ (defined per unit frequency interval). The spectral kinetic energy density $\varepsilon_{th}(\omega)$ of the oscillator can be expressed in terms of the response function, Eq. (3), as

$$\varepsilon_{th}(\omega)d\omega = \tfrac{1}{2}\tilde{m}\,\omega^2\,|\chi(\omega)|^2\,\langle F_{th}^2 \rangle'\,d\omega = \tfrac{1}{2}\tilde{m}\,\langle F_{th}^2 \rangle'\,\frac{d\omega}{4\tilde{m}^2(\omega_0-\omega)^2+\gamma^2}\,. \tag{7}$$

To get the entire mode energy $E_{th}$, one has to integrate over frequency. Evaluating the required integral yields:

$$\int_0^{+\infty}\frac{d\omega}{4\tilde{m}^2(\omega_0-\omega)^2+\gamma^2} = \frac{1}{2\tilde{m}\gamma}\arctan\left(\frac{2\tilde{m}\omega}{\gamma}\right)\Bigg|_{-\omega_0}^{+\infty} \approx \frac{\pi}{2\tilde{m}\gamma}\,, \tag{8}$$

Absolute measurement of thermal noise in a resonant short-range force experiment where exact equality is obtained when the lower integration limit is replaced by $-\infty$. The relative change of the integral due to this is about $10^{-5}$. Then, from Eq.(7),

$$E_{th} = \int_0^{+\infty} \varepsilon_{th}(f)\,df = \frac{1}{2\pi}\int_0^{+\infty} \varepsilon_{th}(\omega)\,d\omega = \frac{1}{8\gamma}\langle F_{th}^2 \rangle' \quad . \tag{9}$$

Setting $E_{th} = \tfrac{1}{2}k_B T$, as prescribed by the equipartition theorem, we arrive at

$$\langle F_{th}^2 \rangle' = 4 k_B T \gamma \quad . \tag{10}$$

This is the same as the mean square random driving force that is stipulated by the Fluctuation-Dissipation theorem by Callen and Greene [1, 29]. The electric analog of this expression was derived much earlier from first principles by Nyquist [30]. It is interesting to note that in the present context, this expression actually *follows* from the equipartition theorem.

### 3.3 Kinetic energy measurement

The average kinetic energy $E_{\text{exp}}$ of the AST mode of the detector can be expressed as:

$$E_{\text{exp}} = \tfrac{1}{2}\tilde{m}\omega_0^2 \langle \tilde{z}_T^2 \rangle \quad , \tag{11}$$

where $\tilde{m}$ is the modal mass (section 4.3) and $\langle \tilde{z}_T^2 \rangle$ is the mean-square displacement of the detector at point $\tilde{P}$ (section 3.1). Capacitive probes near the point $\tilde{P}$ (section 2.4.1) produce a signal that is amplified by the electronics to obtain a mean-square voltage $\langle V_T^2 \rangle$. To eliminate the poorly known transducer efficiency, filter transmittances and amplifier gains, the apparatus is calibrated by observing the voltage response $V_D$ to a displacement amplitude $\tilde{z}_D$ of the oscillator which is imposed by a *known* driving force with amplitude $\tilde{F}$, acting in the $z$-direction at point $\tilde{P}$. This calibration procedure is described in section 4.4.

The ratio $V_D / \tilde{z}_D$ is called the displacement gain $G_z$. Amplification and filter settings are the same for the measurement of the thermal noise and that of the calibration amplitude, and therefore

Absolute measurement of thermal noise in a resonant short-range force experiment

$$G_z = \frac{V_D}{\tilde{z}_D} = \sqrt{\frac{\langle V_T^2 \rangle}{\langle \tilde{z}_T^2 \rangle}} \; . \tag{12}$$

Combining Eqs. (5), (11) and (12), one obtains for the average kinetic energy $E_{exp}$ of the oscillator in terms of measurable quantities:

$$E_{\exp} = \frac{\langle V_T^2 \rangle}{V_D^2} \frac{Q^2 \, \tilde{F}_D{}^2}{2\tilde{m}\omega_0^2} \; . \tag{13}$$

The quantities on the right-hand side of this equation are all measureable or can be modeled. We describe below in detail how this is achieved.

## 4   Experiment
### 4.1   *Quality factor and resonance frequency*

The mechanical quality factor $Q$ is deduced from the decay of the free oscillation of the AST mode. Before exciting the resonance, an approximate value $f_0'$ of the resonance frequency is obtained by observing the thermal noise peak on a spectrum analyzer.[7] The detector is then driven by a small voltage (20 mV) at frequency $f_0'$, applied directly to the detector. In the presence of the probe bias voltage[8], this results in a resonant electrostatic force. During excitation, the transducer amplitude is monitored with the LIA reference frequency set to $f_0'$.

After typically 15 s, the detector is switched to ground and the ring-down is observed with an off-resonance reference frequency ($f_{LIA} = f_0' + 2\,\text{Hz}$). Figure 5 shows the response of the LIA to the free, damped oscillations of the detector. The observed oscillations are the ~2 Hz beats between the actual detector frequency and the reference frequency $f_{LIA}$. The solid line is the result of a least-squares fit to a decaying sine wave:

---

[7] The $Q$ could also be determined from the width of the resonance on the spectrum analyzer, however, the integration times necessary for similar precision would be much longer than the ~ 5 s oscillator ring-down time.

[8] While the symmetric arrangement of the probes is not optimal for *excitation* of the torsional mode, any slight asymmetry in the individual probe gaps results in some of the applied electrostatic signal coupling into this mode.

Absolute measurement of thermal noise in a resonant short-range force experiment

$$V = \overline{V} + V_0 e^{-\frac{t}{\tau}} \cos(2\pi \Delta f_b t + \varphi_b) \quad , \tag{14}$$

where the offset $\overline{V}$, amplitude $V_0$, decay time $\tau$, beat frequency $\Delta f_b$, and phase shift $\varphi_b$ are adjusted parameters. The precise resonance frequency $f_0$ is derived from $f_0 = f_{LIA} - \Delta f_b$ and the mechanical quality factor from $Q = \pi f_0 \tau$.

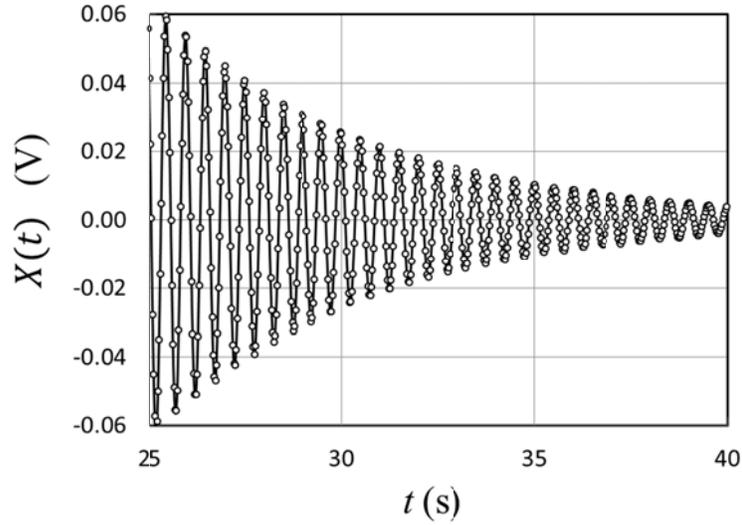

Figure 5. Oscillator ring-down waveform used to extract the value for the mechanical quality factor $Q$. The measured decay time, $\tau$, is 5.41 s.

The following results are averages and standard deviations for a set of four measurements each:

$$f_0 = (1223.395 \pm 0.002) \text{ Hz}, \tag{15}$$

and
$$Q = 20800 \pm 100 \quad . \tag{16}$$

4.2    Observation of thermal noise motion

4.2.1    *Data acquisition.* The spectral density of the random thermal motion of the detector (Eq. (7)) follows a Lorentz distribution, centered at $f_0$ with a width of $\Delta f = f_0/Q \sim 50$ mHz. This narrow peak is superimposed on a flat electronic noise background.[9] The LIA essentially shifts this spectrum by an amount equal to the reference frequency, $f_{LIA}$. Setting $f_{LIA} = f_0$, thus centers the resonance at $f = 0$. The LIA also provides an adjustable low-pass filter with a noise-equivalent bandwidth $B_{LIA}$. Since the spectrum is concentrated near $f = 0$, mostly low frequencies contribute and so the output signals (the quadrature channels $X$ and $Y$) are slowly varying, random voltages. Both of these are sampled at regular intervals $\Delta t$, resulting in a discrete time series $Z_n = X_n + iY_n$ with $n = 1…N$. Complex notation is used for convenience.

The stochastic nature of thermal noise justifies the hypothesis that the processes $X_n$ and $Y_n$ are drawn from a normal distribution and that they are independent, so that their cross-covariance vanishes. It is further assumed that $Z_n$ is stationary, meaning that a statistic derived from the sample is not a function of time. This may be assumed since the measurement time is much longer than the decay time $\tau$ of transients.

As an example, distributions of observed $X_n$ voltages are shown in figure 6. The wider of the two distributions is acquired when the reference frequency is set to $f_{LIA} = f_0$ and the measurement bandwidth is set to $B = 0.5$ Hz, encompassing most of the resonance. To assess the electronic noise contribution, the measurement band is shifted off resonance by setting $f_{LIA} = f_0 - 3\text{Hz}$, which yields the second, narrow distribution. The difference between the two distributions is due to the thermal noise motion. The normality of the observed distribution is confirmed by inspecting an overlay of the cumulative density distribution of the data with that of the normal distribution. In view of the difficulties with normality tests of correlated data, a quantitative distribution analysis has been deferred.

---

[9] The electronic noise attenuates by a nearly constant -1.3 dB/oct over the range from 300 Hz to 6.0 kHz. Thus the fractional change over the width of the resonance is only $10^{-5}$, or $10^{-3}$ for the extended range (5 Hz) analyzed in section 4.2.4.

Absolute measurement of thermal noise in a resonant short-range force experiment

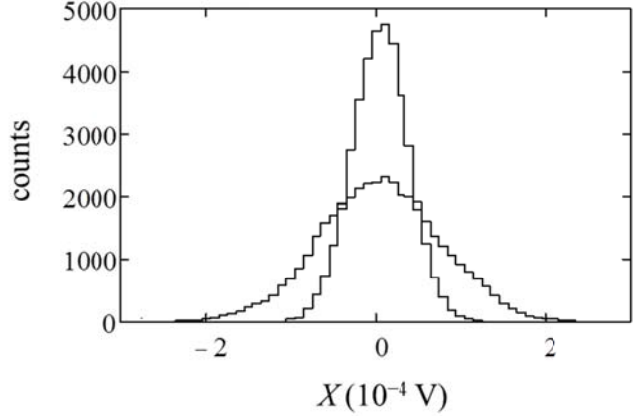

Figure 6. Typical distribution of $X_n$ voltages from the LIA. The narrower distribution is obtained with the LIA tuned 3 Hz below the oscillator resonance and results from electronics noise. The wider distribution, obtained with $f_{LIA} = f_0$, illustrates the additional contribution from the detector motion. There are $N = 40000$ data points in each sample. The bandwidth is set to $B = 0.5$ Hz and the probe bias to $u_p = 108$ V (section 4.2.3).

*4.2.2 Statistical analysis.* The quantity of interest is the noise power of the oscillator within a frequency band $B$, centered around $f_{LIA}$. Within a calibration factor, this equals the mean-square $\langle Z_n^2 \rangle$ of the time series, which in turn equals the variance $\sigma_Z^2$, which can be estimated from the data sample by

$$\hat{\sigma}_Z^2 = \frac{1}{N} \sum_{n=1}^{N} Z_n^* Z_n - |\hat{\mu}_Z|^2 \ , \tag{17}$$

where the caret indicates an estimate from the sample, and the asterisk denotes complex conjugation. The expectation value of the mean equals zero but its sample value

$$\hat{\mu}_Z = \frac{1}{N} \sum_{n=1}^{N} Z_n \tag{18}$$

may contribute to Eq.(17).

Absolute measurement of thermal noise in a resonant short-range force experiment

The uncertainty of $\langle Z_n^2 \rangle$ is the square root of the variance of the variance, $\text{var}(\hat{\sigma}_Z^2)$. For uncorrelated data, $\text{var}(\hat{\sigma}_Z^2)$ is proportional to $1/N$. However, depending on the sampling interval $\Delta t$, the elements of the series $Z_n$ may be correlated since the rate of change of the signal is limited when mostly low frequencies are contributing. The mathematical treatment of correlated time series can be found in standard texts (e.g., [31, 32]). The appropriate expression for $\text{var}(\hat{\sigma}_Z^2)$ turns out to be

$$\text{var}(\hat{\sigma}_Z^2) = \frac{\hat{\sigma}_Z^4}{N}\left[1 + 2\sum_{h=1}^{N-1}\left(1 - \frac{h}{N}\right)\text{Re}(\rho_Z(h))^2\right]. \qquad (19)$$

The second term in the angular brackets represents the effect of correlation, which is measured by the autocorrelation function $\rho_z(h)$. The latter can be estimated from the sample by

$$\hat{\rho}_Z(h) = \frac{1}{N\hat{\sigma}_Z^2}\sum_{n=1}^{N}\left[Z_n^* Z_{n+h} - |\hat{\mu}_Z|^2\right]. \qquad (20)$$

The shift $h$ is known as 'lag', and, obviously, $\hat{\rho}_z(0) = 1$. For stationary processes, $\hat{\rho}_z(h)$ decreases with $h$ and eventually is dominated by statistics. To avoid overestimating the correlation term in Eq. (19), the series may be truncated at that point. Alternatively, an autoregressive model [33] may be fit to the first $p$ values of $\hat{\rho}_z(h \leq p)$, and the model prediction of $\rho_z(h)$ used to complete the series in Eq.(19). The value of $p$ is chosen such that the predicted $\rho_z(h)$ does not change when this parameter is increased. In the present case, this fairly involved procedure has been found to add at most a few percent to the correlation term, and will not be further discussed here.

Increasing the number $N$ of samples in a data set by lowering the sample time $\Delta t$ would decrease $\text{var}(\hat{\sigma}_Z^2)$, but is counteracted by the correlation term in Eq. (19). When $\Delta t$ is small enough $\text{var}(\hat{\sigma}_Z^2)$ is completely independent of $\Delta t$. The present $\Delta t = 0.1$ s is well below this limit.

Absolute measurement of thermal noise in a resonant short-range force experiment

*4.2.3 Dependence on probe bias.* The observed voltage signal is proportional to the bias $u_p$ between the detector and the capacitive transducer probes (section 2.4.1). Thus, the sensitivity to detector oscillations may, in principle, be increased by increasing $u_p$. There are practical limits, however, since the probe field could be the cause of instabilities.

To decide on a stable operating point for the particular probe gaps set (~ 100 µm), the oscillator noise is measured as a function of probe bias $u_p$. Results are shown in figure 7. Data are acquired by evaluating the variance of the recorded time series $Z_n$, Eq.(17), and *its* variance, Eq.(19). One needs to take into account that the probe bias field exerts a force on the oscillator proportional to the inverse square of the gap, which slightly modifies the restoring potential and affects the resonance frequency (by about 0.5 Hz over the range of the measurement). For this reason, $f_0$ is determined with a spectrum analyzer (to within 15 mHz) after to each setting of $u_p$.

For all subsequent measurements, as well as the $Q$ and $f_0$ measurements in section 4.1, the setting $u_p$ =108 V is selected. This value is well within the linear range and still a factor of 2 above the amplifier noise (open circles).

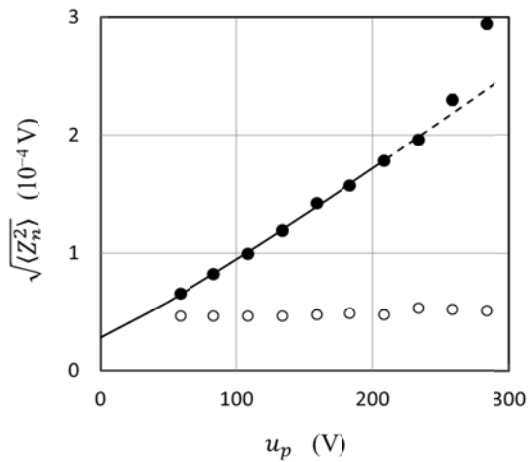

Figure 7. Observed rms noise signal as a function of the transducer bias $u_p$. The bandwidth is set to $B = 0.5$ Hz. Full circles: the LIA is tuned to the resonance ($f_{LI} = f_0$). The solid line is a fit to a power-law function (the exponent is 1.14), excluding data at $u_p > 250$ V, where there is a clear departure from linearity. The dashed line is an extension of the fit curve in that range. Open circles: electronics noise, measured off-resonance ($f_{LI} = f_0 - 2$Hz). The two data points at $u_p = 108$ V correspond to the two distributions shown in figure 6.

Absolute measurement of thermal noise in a resonant short-range force experiment

*4.2.4   Thermal noise measurement.* The spectral density of the signal after electronic processing consists of a constant electronic noise background with a resonance peak at $f_0$. The data are acquired by evaluating the variance, Eq. (17), and its variance, Eq. (19) of the recorded time series $Z_n$.

Let $\langle Z(B)^2 \rangle$ represent the measured noise power within a frequency band $B$, centered at $f_0$. The bandwidth is controlled by the LIA. Data are acquired as a function of $B$, either 'on-resonance', $\langle Z_\alpha(B)^2 \rangle$, or 'off-resonance', $\langle Z_\beta(B)^2 \rangle$. The duration of a measurement is commensurate with the bandwidth and ranges from a few hundred seconds to 16000 s. The sampling interval is $\Delta t = 0.1$ s.

For the off-resonance measurement, the LIA reference frequency is set to $f_{LIA} = f_0 - 2\text{Hz}$. In the case of white background noise, the measured power is proportional to the bandwidth:

$$\langle Z_\beta(B)^2 \rangle = S_\beta \cdot B \quad , \tag{21}$$

where $S_\beta$ is the spectral density of the white electronic background noise (open symbols in figure 8).

Measurements on resonance, $\langle Z_\beta(B)^2 \rangle$ (solid symbols), exhibit the additional contribution due to the motion of the oscillator,

$$\langle Z_\alpha(B)^2 \rangle = S_\beta \cdot B + \langle V_T^2 \rangle \left[ \frac{2}{\pi} \arctan\left( \frac{QB}{f_0} \right) \right] \quad . \tag{22}$$

The term in angular brackets is the fraction of a Lorentz distribution between $f_0 - B/2$ and $f_0 + B/2$ (see Eq.(8)). For large $B$, this term tends to 1, and $\langle V_T^2 \rangle$ represents the offset between the parallel parts of the upper and lower curve. Thus, $\langle V_T^2 \rangle$ equals the total noise power due to the resonant detector motion. These quantities are determined from a simultaneous fit of Eqs. (21) and (22) to the on- and off-resonance data, resulting in

$$\langle V_T^2 \rangle = (7.9 \pm 0.3) \cdot 10^{-9} \; \text{V}^2 \tag{23}$$

and $\qquad\qquad\qquad S_\beta = (4.25 \pm 0.007) \cdot 10^{-9} \; \text{V}^2/\text{Hz.} \tag{24}$

Absolute measurement of thermal noise in a resonant short-range force experiment

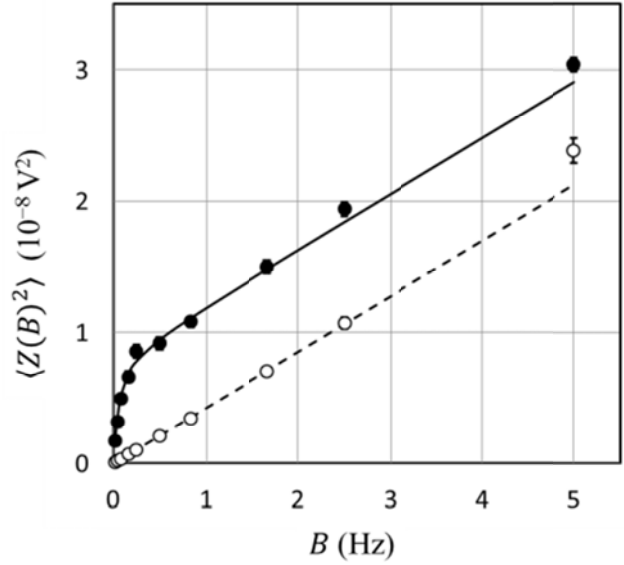

Figure 8. Noise power as a function of measurement bandwidth $B$. The curves represent a simultaneous fit with Eqs. (21) and (22). The probe bias is $u_p$ = 108 V. The off-resonance point at $B$ = 5 Hz is not used in the fit because in this case, the measurement bandwidth includes part of the tail of the resonance.

*4.2.5 Spectral analysis.* Spectral analysis is the estimation of the spectral density of a random signal from a sequence $Z_n$ of time samples. The purpose of applying spectral analysis to our data is to show that the spectrum is consistent with a Lorentz distribution with an area that agrees with the thermal noise power of (Eq. (23)).

The spectral power density distribution $S(f)$ describes the power per frequency interval (in V$^2$/Hz) as a function of frequency $f$, up to a maximum $f_{max} = 1/2\Delta t$, given by Nyquist's theorem. The Wiener-Khintchin theorem relates the spectral distribution of a time series to the auto-correlation function (see standard texts on time series, e.g., [31, 32]).

The estimate of $S(f)$ from the data can be written as

$$\hat{S}(f) = 2\Delta t \hat{\sigma}_z^2 \left\{ 1 + 2 \sum_{h=1}^{H} \lambda(h) \operatorname{Re} \hat{\rho}_z(h) \cos(2\pi f \cdot \Delta t \cdot h) \right\}, \qquad (25)$$

Absolute measurement of thermal noise in a resonant short-range force experiment where $\hat{\rho}_z(h)$ is given by Eq. (20). Obviously, the integral of $\hat{S}(f)$ from 0 to $f_{max}$ equals the variance $\hat{\sigma}_Z^2$, i.e., the total power. The sum is terminated at some arbitrary $H$, to suppress spurious contributions, with $H = 2\sqrt{N}$ recommended as a first choice. Furthermore, a 'window' $\lambda_h$ is introduced to modify the terms near the upper end of the sum. Options for choosing a window are discussed in [31]); arbitrarily, a Tukey window is used here:

$$\lambda(h) = \frac{1}{2}\left(1 + \cos\left(\frac{h \cdot \pi}{H}\right)\right). \tag{26}$$

To apply this procedure, dedicated data are acquired with a reference frequency $f_{LIA} = f_0 + 0.20$ Hz. This moves the resonance peak away from $f = 0$. The bandwidth $B_{LIA} = 0.83$ Hz is chosen to still fully cover the resonance. The measured series contains $N = 20000$ data points; $H = 282$ is picked as a cut-off. The solid line in figure 9 is obtained by evaluating Eq. (25). The dotted line is a Lorentz distribution consistent with the $Q$ value listed in Eq. (16), reproducing the measured resonance very well.

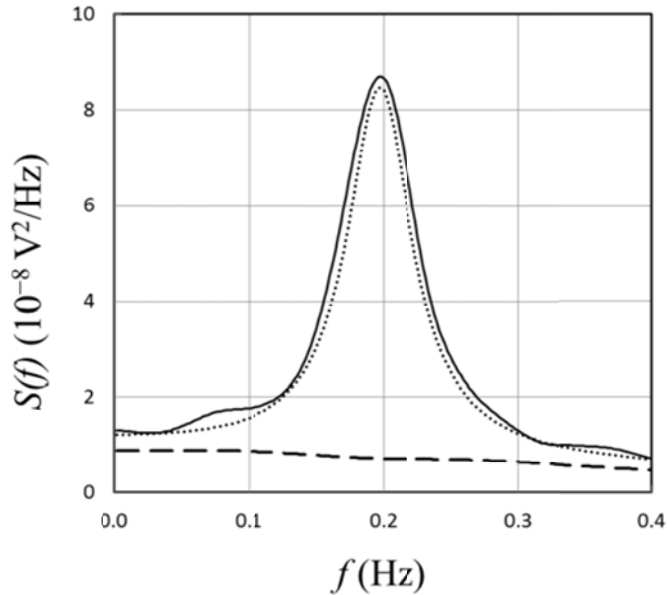

Figure 9. Spectral power density distribution $S(f)$, estimated from the data (solid line). For comparison, a Lorentz distribution with the $Q$ value of Eq.(16) is shown as a dotted line. The dashed line indicates the off-resonance background.

Absolute measurement of thermal noise in a resonant short-range force experiment

The electronic noise background is determined by setting $f_{LIA} = f_0 - 3$ Hz (dashed line). The non-flatness of the background in figure 9 comes from the fact that the spectral analysis packs all frequencies in the range from 0 to $f_{max}$, so there is wrap-around. The area under the background-corrected peak is consistent with the thermal noise power of Eq. (23).

*4.3  Mode shape and modal mass*

The Structural Mechanics Module of the finite-element code COMSOL is used to model the motion of the detector. The input required includes the geometry of the detector (figure 3) and the density $\rho_W$, Young's modulus, and Poisson ratio of tungsten. The base plate (B) is held fixed. The division into mesh elements is shown in figure 3.

The code establishes the eigenmodes of the system and, for each, calculates the resonance frequency and the displacement amplitude at all nodes of the mesh (for some arbitrary excitation at the resonance frequency). Information on the stress is also available but is not used in this context. The AST mode with which we are concerned here corresponds to the 5th eigenmode, in ascending order of frequency, detected by the code. For the present oscillator, the calculated resonance frequency equals 1227 Hz (to be compared with Eq. (15)).

The *mode shape* $\xi(x,y)$ (section 3.1) is the calculated displacement amplitude, normalized some arbitrary point $P_0$, such that $\xi(x_0, y_0) = 1000$. For ease of use, a rectangular grid is defined on the surface of the detector and the mode shape $\xi(x_i, y_k)$ is calculated at the corners of this grid by interpolating the exported displacement values. Since all dimensions of the detector are multiples of 0.025", a grid size $\Delta s = 0.635$ mm is chosen, putting grid points on all corners and edges of the detector. On account of the axial symmetry, only half of the detector is covered. The mode shape value at the reference point $\tilde{P}$ equals $\tilde{\xi} = -273$.

The *modal mass* $\tilde{m}$ is associated with a given reference point $\tilde{P}$ and is *defined* by the average kinetic energy $E_{kin}$ of the detector in terms of the displacement $\tilde{z}$ at that point $\tilde{P}$,

$$E_{kin} \equiv \tfrac{1}{2}\tilde{m}\omega^2 \langle \tilde{z}^2 \rangle \quad . \tag{27}$$

The mode shape relates the motion at $\tilde{P}$ to the detector motion anywhere else (Eq.(1)), and it is easy to see that

Absolute measurement of thermal noise in a resonant short-range force experiment

$$\tilde{m} = \frac{\Delta m}{\tilde{\xi}^2} 2\sum_{i,k} w_{ik}\, \xi^2(x_i, y_k) \ . \tag{28}$$

The sum extends over all grid points (covering half of the detector) and $w_{ik} = 1$, except for grid points on the edges ($w = \frac{1}{2}$), or at the corners ($w = \frac{1}{4}$). The mass of a grid element equals $\Delta m = \rho_W d_W \Delta s^2$. Combining information on the tungsten raw material, literature values on $\rho_W$ and measurements of the weight and dimensions of the detector, the area density is $\rho_W d_W = 3.55$ mg/mm$^2$ with an estimated 5% error. Since the mode shape neither depends on the density nor the elastic constants, its contribution to the uncertainty of $\tilde{m}$ is neglected. The modal mass is then

$$\tilde{m} = (0.93 \pm 0.05) \text{ g}. \tag{29}$$

COMSOL can be used to calculate the kinetic energy directly (as a derived quantity). Inserting the corresponding value into Eq. (27) leads to 0.944 g for the modal mass, which checks with Eq. (29), as expected.

*4.4  Calibration*

*4.4.1  Driving force.* A calibration of the experiment requires a force acting on the oscillator which is known *a priori*. To generate such a force electrostatically, the source and detector plates are used as electrodes in a parallel-plate capacitor. To this effect, the source plate is positioned parallel to the detector oscillator, separated by a gap $z_g$, in such a way that it overlaps exactly half of the front rectangle F in figure 3. With the detector at ground, a voltage $U_D(t) = U_{D0} \sin(\omega_D t)$ is applied to the source, where $\omega_D = 2\pi f_D$ is the driving frequency. (The source is present in the same position for both thermal and calibration measurements; the voltage is applied only during calibration.) Assuming that the electrical field is uniform between the plates and zero elsewhere, the force per unit area is

$$\Phi(t) = \frac{\varepsilon_0}{2z_g^2} U_D(t)^2 = \Phi_{const} - \frac{\varepsilon_0}{4z_g^2} U_{D0}^2 \sin(2\omega_D t) \ , \tag{30}$$

where only the time-dependent part is of interest. The detector is excited on resonance when the driving frequency equals half the resonance frequency.

To better insure that the calibration force be calculable, the test mass gap $z_g$ is chosen to be in a range 10 times greater than the largest surface height variation (about 10 microns) measured during the test mass leveling procedure, but less than 10% of the length of the shortest edge (5 mm) of either test mass. To ensure that the calibration signal is dominated by the applied electrostatic force, the signal is studied as a function of both the gap and the applied voltage $U_D(t)$ to establish the ranges exhibiting the desired $(U_{D0}/z_g)^2$ behavior.

To maintain a stationary process, the work done on the detector by the driving force must make up for the energy lost by friction. This requirement makes it possible to convert a distributed electrostatic force into an equivalent point force acting in the $z$-direction at the reference point $\tilde{P}$. It is straightforward to show that in this case the amplitude of this equivalent driving force is

$$\tilde{F}_D = \frac{\varepsilon_0}{4z_g^2} U_{D0}^2 \frac{\Delta s^2}{\tilde{\xi}} \sum_{i,k} w_{ik}\, \xi(x_i, y_k) \ . \tag{31}$$

Here, $\xi(x,y)$ refers to the mode shape (section 4.3). The sum extends over all grid points in the overlap region and $w_{ik} = 1$, except for grid points on the edges ($w = \frac{1}{2}$), or at the corners ($w = \frac{1}{4}$). The area of a grid element equals $\Delta s^2 = 0.4032$ mm$^2$, the gap size is $z_g = 0.381$ mm, and the voltage amplitude $U_{D0} = 0.8$ V. The result for the inferred driving force amplitude is then

$$\tilde{F}_D = (4.9 \pm 0.4)\cdot 10^{-10}\ \mathrm{N}\ . \tag{32}$$

The quoted uncertainty has contributions from the gap size (~8%) and the relative source-detector positioning, affecting the overlap area (~2%).

The effect of the fringe fields between the two plates on the force between them has been studied using the AC/DC Module of the finite-element code COMSOL. To this effect, the $z$-component of the Maxwell surface stress tensor on the surface of the detector is evaluated. The resulting, position-dependent area force density, weighted with the relative mode shape, is then integrated over the detector surface. The result is the same as Eq.(32) to within 0.1%; in other words, the driving force is not modified significantly by the fringe field.

*4.4.2 Detector response.* The signal from the pick-up near point $\tilde{P}$ of the driven oscillator is amplified, filtered and processed by the LIA. The reference signal to the LIA is chosen to be in phase with the

Absolute measurement of thermal noise in a resonant short-range force experiment driving signal at frequency $f_D$. For a measurement sequence, the driving frequency is stepped across the resonance, while the quadrature outputs $X(f_D)$ and $Y(f_D)$ of the LIA are recorded.

The response of the detector to the force described above is shown in figure 10. Different symbols correspond to three consecutive sweeps with different LIA bandwidths.

As the frequency is varied across the resonance, the response function $\chi(\omega)$, Eq.(3), traces out a circle in the complex plane, which tends towards the origin for frequencies far from resonance. Consequently, the same holds true for the LIA output $Z(f_D) = X(f_D) + iY(f_D)$. The diameter of this circle equals the resonance amplitude $|Z(f_0)|$. The circle parameters are adjusted to the data, minimizing the sum of the closest distances of the data to the circle. From this follows the circle diameter and thus the voltage amplitude $V_D$ associated with the driving force from Eq.(32)

$$V_D = |Z(f_0)| = (61.95 \pm 0.05)\ \text{mV}. \qquad (33)$$

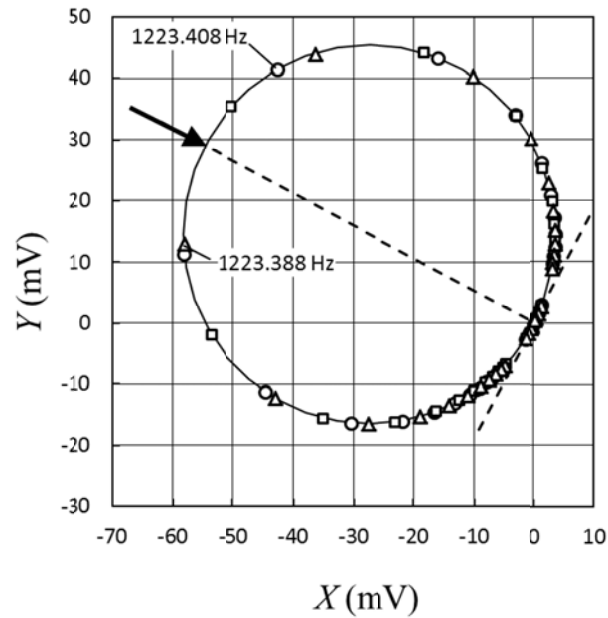

Figure 10. Response of the oscillator to the driving force as a function of driving frequency. The quadrature output $Y$ is plotted versus the in-phase output $X$ of the LIA. Close to resonance (indicated by an arrow), $f_D$ is changed in steps of 20 mHz. Circles, squares and triangles mark data taken with LIA bandwidths of 0.041 Hz, 0.250 Hz and 1.25 Hz. Two points are labeled with their actual frequencies, as example. The solid line is the result of a least-squares fit to a circle.

Absolute measurement of thermal noise in a resonant short-range force experiment

## 5 Conclusions

Inserting the numerical results given in Eqs. (15), (16), (23), (29), (32), and (33) into Eq. (13) yields as the final result for the measured average kinetic energy of the detector oscillator due to thermal motion:

$$E_{exp} = (2.0 \pm 0.3) \cdot 10^{-21} \text{ J} . \qquad (34)$$

The uncertainty is dominated by the calibration driving force and by the calculation of the modal mass. The measured $E_{exp}$ is consistent with the thermal kinetic energy $\frac{1}{2}k_B T$ (Eq. (6)) as expected from the equipartition theorem. This agreement confirms that the observed random motion of the detector is due to the fundamental coupling of the device to the temperature bath of the environment. This conclusion is also supported by the stochastic properties of the observed time series when measuring the thermal noise. By extension, it confirms the effectiveness of the vibration isolation system against environmental factors such as seismic and room vibrations.

This experiment is an illustration of the remarkable sensitivity that can be attained with high-$Q$ mechanical oscillators. After all, the measured energy in Eq. (34) is three times smaller than the kinetic energy of a single molecule at the same temperature. The rms displacement of the reference point $\tilde{P}$ due to thermal motion is about 300 fm, and the displacement gain, Eq.(12), which relates the processed signal to the displacement at point $\tilde{P}$ is

$$G_z = (3.3 \pm 0.3) \cdot 10^8 \text{ V/m} . \qquad (35)$$

The equivalent thermal noise force, Eq. (10), provides a means to estimate the sensitivity of the present experiment to a driving force of arbitrary origin. Averaging over the duration of a typical run, say $\tau = 24$ hours, one finds that a force amplitude (at point $\tilde{P}$) of $F = \sqrt{4kT\gamma/\tau} = 8$ fN would become visible. For a force at the point $P_0$ in figure 3 (the quantity reported in [20]), the mode shape function suggests a sensitivity of

$$8\text{fN} \times \frac{\tilde{\xi}}{\xi(x_0, y_0)} = 2 \text{ fN}, \qquad (36)$$

essentially the same result measured in [20].

A more illustrative quantity is the sensitivity to a force amplitude averaged over the area of detector rectangle F on one side of the torsion axis, which better represents the region over which a driving force is typically applied. The result for the average $\xi$ (in both the current and 2003 finite element models) is very nearly 0.5 μm, for a force sensitivity of about 4 fN. For the test masses in figure 2, with a minimum gap of 50 μm, this is larger than the edge-effect remnants of the resonant Newtonian $1/r^2$ force (1.7 fN, calculated by numerical integration). For the same geometry, it is equivalent to a Yukawa-type force of about 3000 times gravitational strength and a range of 10 μm, which is about an order of magnitude smaller than the current experimental limits in that range [16, 17]. The potential improvement in sensitivity at this range derives from the rigid design possible at 1 kHz, which permits operation with even smaller test mass separations than previously attained with this technique. This assumes, of course, that systematic backgrounds can be controlled with the thinner shield.

From the information presented in this paper, it is obvious that the sensitivity of the present apparatus, when used as a short-range force search experiment, is limited by thermal noise. On account of the higher frequency relative to torsion pendulum experiments such as those at HUST and UW, the absolute sensitivity is several orders of magnitude less than those experiments when operating near the thermal noise limit. When the thermal noise limit has been reached in a force search, the absolute sensitivity at all ranges can be improved by an order of magnitude or more by cooling the experiment to cryogenic temperatures. An additional improvement is expected from an increase in the *Q*-value. Such an increase has been observed in similar annealed tungsten oscillators when their temperature was lowered [27].

**Acknowledgements**

The authors thank B. Bartram and E. Luther of the MST-6 group at the Los Alamos National Laboratory for performing the high-temperature vacuum annealing of the oscillator prototypes. This work is supported by National Science Foundation grants PHY-1207656 and DMS-1206405, the Indiana University Collaborative Research Grant (IUCRG) program, and the Indiana University Center for Spacetime Symmetries (IUCSS).

Absolute measurement of thermal noise in a resonant short-range force experiment